\documentclass[aps,prd,showpacs]{revtex4}
\usepackage{amssymb}
\usepackage{graphicx}

\newcommand{\md}{{\mathrm{d}}}

\begin{document}

\title{
Classical versus quantum evolution for a universe with a positive \\ cosmological constant
}

\author{David Brizuela}
\email{david.brizuela@ehu.eus}
\pacs{03.65.-w, 03.65.Sq, 98.80.Qc}

\affiliation{Fisika Teorikoa eta Zientziaren Historia Saila, UPV/EHU, 644 P.K., 48080 Bilbao, Spain}
\affiliation{Institut f\"ur Theoretische Physik, Universit\"at zu K\"oln, Z\"ulpicher Stra{\ss}e 77, 50937 K\"oln, Germany}

\begin{abstract}
A homogeneous and isotropic cosmological model with a positive cosmological constant is considered.
The matter sector is given by a massless scalar field, which can be used as an internal time
to deparametrize the theory. The idea is to study and compare the evolutions of a quantum and a
classical probability distribution by performing a decomposition of both distributions
in their corresponding moments. For the numerical analysis
an initial peaked Gaussian state in the volume will be chosen.
Furthermore, in order to check the robustness of certain results,
as initial state both a slightly deformed Gaussian, as well as another completely different
state, will also be studied.
Differences and similarities between classical and quantum
moments are pointed out. In particular, for a subset of moments classical and quantum
evolutions are quite similar, but certain variables show remarkable differences.
\end{abstract}

\maketitle

\section{Introduction}

During the last years an intensive effort is being made to construct
effective theories (understood as a systematic framework that provides the
classical equations of motion plus certain quantum corrections)
for quantum cosmology \cite{Boj12}. The main motivation
is to obtain testable results in scenarios where the complete knowledge
of the underlying fundamental quantum gravity theory is not necessary,
in such a way that quantum cosmology becomes an empirical science.
In this respect, there are already several proposals and approaches in order to
find quantum-gravity corrections to the anisotropy spectrum of the
cosmic microwave background (see for instance \cite{BCS11, KiKr12, BEKK13, KTV13, CFM14, AAN13, BCG14}).

In the particular case of loop quantum cosmology \cite{loops},
there are three key ingredients that one should consider in order to construct
such an effective theory: \emph{holonomy} corrections, \emph{inverse-triad} corrections,
and \emph{quantum-dynamical} corrections. The origin of
the first two corrections lies in the variables that are used in this specific
theory: holonomies of connections and fluxes of spatial triads.
Holonomy corrections appear in a process of regularization, and they take the form of higher powers of the connection
that amend the classical Hamiltonian \cite{RS94}. On the other hand, the
inverse-triad corrections are produced because, in order to
avoid infinities, the inverse of a given triad is replaced by the
Poisson bracket between the corresponding triad and certain holonomy,
which constitutes a classical identity \cite{Thi98}. This procedure prevents,
at the quantum level, the operator associated with the inverse of the triad
from diverging, even when the triad itself tends to zero.
The latter (quantum-dynamical) corrections are not specific to a loop quantization and arises
due to the distributional character of quantum mechanics and the noncommutativity of
the basic operators. In the present paper we will focus
on these corrections, so let us analyze their origin in more detail.

In the procedure of quantization each classical degree of freedom
is replaced by an infinite set of quantum degrees of freedom, usually
described by a probability distribution (wave function). Another way to parametrize these degrees
of freedom is by decomposing the wave function in its infinite set
of moments.
These moments then appear in the classical Hamilton equations as quantum corrections.
Nonetheless, this distributional character is not specific
of the quantum theory. In fact, only ideally, classical mechanics is non-distributional.
When the initial conditions are not known with infinite precision, this uncertainty
can be described by an initial probability distribution, and it is then necessary to
consider the evolution of such a distribution on the classical phase space.
As explicitly shown in \cite{BM98, Bri14}, the evolution of a classical distribution
can also be described in terms of its moments. And, interestingly, it turns out that
the equations of motion for these classical moments can be obtained from the equations
of motion for the quantum moments just by imposing a vanishing value of the Planck constant. This is
a neat classical limit of a quantum theory, which implements the idea
presented in \cite{BYZ94} that the limit of a quantum theory is not a unique
orbit on phase space, but an ensemble of classical orbits.
Note that the mentioned classical and quantum probability distributions are defined on different
spaces. Thus, their decomposition in moments allows us to compare their evolution.
In addition, moments represent observable quantities that could be, in principle,
experimentally measured.

Therefore, one can distinguish two different origins
of quantum-dynamical corrections. On the one hand, the fact that an extensive (as opposed to a Dirac delta)
distribution needs to be considered makes the presence of moments in the equations of motion unavoidable.
Nonetheless, these kind of terms are also present in the evolution of a classical distribution
and thus they are not genuinely quantum. For instance, as in the quantum case, they generically prevent the
centroid of a classical distribution (the expectation value of the position and momentum)
from following a classical trajectory on the phase space. On the other hand, the \emph{purely quantum}
or \emph{noncommutativity} terms arise due to the noncommutativity of the basic operators.
In the equations of motion they appear as a power series in $\hbar^2$.
Following the terminology of \cite{Bri14}, the first ones will be named \emph{distributional} effects,
whereas the latter ones \emph{purely quantum} effects. It is important to stress that,
if quantum cosmology is to become a testable theory,
one needs to discriminate between purely quantum effects and effects that might appear just
due to different technical or measurement errors, which would imply a classical probability distribution.

Let us briefly mention that this formalism based on a decomposition of the wave function in terms of its moments
was first presented in \cite{BM98} for the case of a particle on a potential. A similar formalism was derived
in \cite{BoSk06, BSS09} for generic Hamiltonians and on a canonical framework, but making use of a different
ordering of the basic variables. Furthermore, this formalism has been adapted to the case when
the dynamics is ruled by a Hamiltonian constraint, as opposed to a Hamiltonian function \cite{BoTs09}.
This is of particular importance in the context of general relativity, where the Hamiltonian is a linear combination
of constraints.
It has also been extensively applied to different models
of quantum cosmology: isotropic models
with a cosmological constant have been studied in \cite{BoTa08, BBH11} whereas,
in the context of a loop quantization, bounce scenarios have been analyzed in \cite{Boj07}.
The problem of time has also been analyzed in \cite{BHT11, HKT12} within this framework.
Finally the classical counterpart of the formalism developed in \cite{BoSk06, BSS09}, for the analysis
of the evolution of a classical distribution in terms of its moments,
was presented in \cite{Bri14} and applied to the case of a particle on a potential in \cite{Bri14b}.

In this paper we will revisit, from the perspective of a classical probability distribution,
the model studied in Ref. \cite{BBH11} for the evolution of quantum moments on a homogeneous and isotropic
universe with a positive cosmological constant. The numerical analysis
of the present model in terms of a wave function is presented in \cite{PaAs12}, both for a geometrodynamical
and a loop quantization. Here moments will be defined in terms of geometrodynamical variables.

Our goal is to find differences and similarities
between the quantum and classical (distributional) evolution of this cosmological model, given the same initial
data for both cases, in order to study whether any moment has some distinctive or characteristic
behavior under either the quantum or classical evolutions. As already commented above, the great advantage of the present formalism is that
it considers the evolution of moments, which are observables, and no mention to an abstract
mathematical object, as the wave function, has to be made. Therefore, choosing the same initial data for
both classical and quantum sectors is straightforward. This kind of question is very difficult to be
posed in terms of a wave function, since one needs to somehow define its classical probability
distribution analog and choose the {\it same} initial data. (This can be done via, for instance,
the Wigner transform but in general it is not positive definite and thus can not be strictly understood as a probability distribution.)
Finally note that the present analysis is different from tying to define a {\it classicalization} of a given quantum system.
This latter is usually addressed by appealing to the WKB limit or different decoherence processes
that annihilate the quantum interference (see, e.g., \cite{Hab90, Kie92, Zur03}).

The rest of the paper is organized as follows. In Sec. \ref{sec_formalism}
the formalism for the evolution of classical and quantum probability distributions
in terms of their corresponding moments is briefly summarized.
Section \ref{sec_cosmology} presents the specific cosmological model under
consideration. In Sec. \ref{sec_numerical} the results obtained with different
numerical implementations are described.
Finally, Sec. \ref{sec_conclusions} discusses the main conclusions.

\section{General formalism}\label{sec_formalism}

Let us assume a quantum mechanical system parametrized by the
conjugate variables $(\hat q, \hat p)$. The quantum moments associated to this system
are given by
\begin{equation}
G^{a,b}:=\langle(\hat p - p)^a \,(\hat q - q)^b\rangle_{\rm Weyl},
\end{equation}
where $p:=\langle\hat p\rangle$, $q:=\langle\hat q\rangle$ and
the subscript Weyl stands for totally symmetric ordering.
The order of a moment $G^{a,b}$ is defined as the sum between its
two indices $(a+b)$. Note that through this decomposition the wave function
that describes the quantum state of the system gets
replaced by its infinite set of moments $G^{a,b}$, which only depend on time.
Moments that correspond to a valid wave function must fulfill certain relations
due to the Schwarz inequalities, see \cite{Bri14} for a systematic derivation
of such relations. The simplest, and probably most important,
example is the Heisenberg uncertainty principle that, with this notation,
takes the following form:
\begin{equation}\label{heisenberg}
(G^{1,1})^2\leq  G^{2,0}G^{0,2} -\frac{\hbar^2}{4}\,.
\end{equation}

The dynamical information of these moments is encoded on an effective Hamiltonian $H_Q$,
which is obtained by performing a Taylor expansion of the expectation value
of the Hamiltonian operator around the centroid:
\begin{eqnarray} \label{HQ}
H_Q(q,p,G^{a,b})&:=&\langle\hat{H}(\hat q, \hat p)\rangle_{\rm Weyl}
=\langle\hat{H}(\hat q-q+q, \hat p -p +p)\rangle_{\rm Weyl}
= \sum_{a=0}^\infty\sum_{b=0}^\infty \frac{1}{a!b!}
\frac{\partial^{a+b} H}{\partial p^a\partial q^b} G^{a,b}\nonumber\\
&=&H(q,p) + \sum_{a+b\geq 2}\frac{1}{a!b!}\frac{\partial^{a+b} H}{\partial p^a\partial q^b} G^{a,b}.
\end{eqnarray}
In order to obtain the evolution equations for the moments $G^{a,b}$, as well as for the expectation values
($q$, $p$), it is enough to compute the Poisson brackets between each of these variables and the
above Hamiltonian $H_Q$. In this way one obtains the following infinite set of ordinary differential
equations:
\begin{eqnarray}\label{eqqq}
\frac{dq}{dt}&=&\frac{\partial H(q,p)}{\partial p}+\sum_{a+b\geq2}\frac{1}{a!b!}\frac{\partial^{a+b+1}H(q,p)}{\partial p^{a+1}\partial q^{b}}G^{a,b},\\
\frac{dp}{dt}&=&-\frac{\partial H(q,p)}{\partial q}-\sum_{a+b\geq2}\frac{1}{a!b!}\frac{\partial^{a+b+1}H(q,p)}{\partial p^{a}\partial q^{b+1}}G^{a,b},
\label{eqpq}\\\label{eqg}
\frac{d G^{a,b}}{dt}&=&\{G^{a,b},H_Q\}
=\sum_{c+d\geq 2} \frac{1}{c!d!}
\frac{\partial^{c+d} H}{\partial p^c\partial q^d} \{G^{a,b},G^{c,d}\},
\end{eqnarray}
which is completely equivalent to the flow of states
generated by the Schr\"odinger equation.
Note that the first two equations are the usual Hamilton equations
plus certain correction terms that depend on the moments. If all moments were
vanishing, these terms would completely disappear. One of the consequences of these terms
is that the centroid of a quantum distribution $(q,p)$ does not follow a \emph{classical
point orbit} (the orbit obtained with an initial Dirac delta distribution, for which
all moments vanish). Even so, this effect is not genuinely quantum since, as will
be shown below, also happens for classical probability distributions.
A closed formula for the Poisson brackets between any two moments, which
has been left indicated in Eq. (\ref{eqg}), can be found in Ref. \cite{BSS09,BBH11}.
For practical purposes, generically one needs to truncate the infinite system by
introducing a cutoff by hand in order to be able to solve it.

On the other hand, let us assume an ensemble on the classical phase space coordinatized
by  the conjugate variables $(\tilde q, \tilde p)$. Such an ensemble will be described by
a probability distribution function $\rho(\tilde q, \tilde p,t)$, which will obey the Liouville
equation. This distribution function defines a natural expectation value operation on
the classical phase space for any function $f(\tilde q,\tilde p)$ in the following way, 
\begin{equation}
\langle f(\tilde q,\tilde p) \rangle_{\rm c} := \int d\tilde q d\tilde p f(\tilde q,\tilde p) \rho(\tilde q,\tilde p,t),
\end{equation}
where the integration should be taken along the domain of the probability distribution.

Following the same procedure as in the quantum case, the classical moments
are then defined as
\begin{equation}
C^{a,b}:=\langle (\tilde p-p)^a (\tilde q-q)^b \rangle_{\rm c},
\end{equation}
where $(q,p)$ are the coordinates of the centroid of the distribution: $q:=\langle\tilde q \rangle_c$
and $p:=\langle\tilde p \rangle_c$. [For notational simplicity, and for the moment being,
the same notation $(q,p)$ will be used for the centroid of both classical and quantum distributions.
Nonetheless, in Subsec. \ref{sec_cosmologymoments} specific notations will be introduced for each
of them, to be used when
the meaning is not clear from the context.] On the contrary to the quantum case, here all variables
commute and, therefore, the ordering inside the expectation value is irrelevant.
The Hamiltonian that describes the evolution of these classical variables is obtained,
as in the quantum case, by expanding the expectation value of the Hamiltonian around
the position of the centroid:
 \begin{eqnarray}
H_{\rm C}(q,p,C^{a,b}):=\langle  H(\tilde q,\tilde p) \rangle_{\rm c}
=H(q,p) + \sum_{a+b\geq2}\frac{1}{a!b!}\frac{\partial^{a+b} H(q,p)}{\partial p^a\partial q^b}C^{a,b}.
\end{eqnarray}
In order to get the evolution equations for the classical expectation values $(q,p)$
and moments $C^{a,b}$ one just needs to compute the Poisson brackets
of different variables with this Hamiltonian. The flow generated by the Liouville equation
for the probability distribution $\rho(\tilde q, \tilde p, t)$ is then equivalent
to the infinite set of equations obtained by this procedure.

It turns out that
the difference between the quantum and classical equations are just terms
that appear in the former equations multiplied by even powers of Planck constant $\hbar$,
and are missing in the classical equations.
Such $\hbar$ factors are present in the quantum system due to the noncommutativity
of the basic operators $\hat q$ and $\hat p$.
In fact, the classical equations of motion for the moments and expectation values
can be obtained from their quantum counterparts by imposing
a vanishing value of the Planck constant. Thus the classical limit of a quantum
system, understood as $\hbar\rightarrow 0$, turns out to be very neat in this formalism. In particular, as can be seen,
such a limit does not give a unique trajectory on the classical phase space but an
ensemble of them.

Due to the properties of the equations mentioned above, it is possible to classify the quantum
effects depending on its origin. On the one hand, \emph{distributional} effects arise because,
 due to the Heisenberg uncertainty relation, all quantum moments can not be vanishing. These
effects are also present in the classical setting when considering the evolution of a spread
probability distribution, for instance in the usual situation where the initial data are not
known with infinite precision. On the other hand, the \emph{noncommutativity}
or \emph{purely quantum} effects appear in the quantum equations as
explicit $\hbar$ terms. As commented above, the origin of such terms lies in
the noncommutativity of the basic operators. On the contrary to distributional effects,
these are not present in the classical setting and are, thus, genuinely quantum.

There are two classes of Hamiltonians that have
very special properties regarding the classical and quantum evolution they generate \cite{Bri14}.
On the one hand, the quantum equations derived from any harmonic Hamiltonian,
which are at most quadratic on the basic variables, do not contain any $\hbar$ term.
Therefore, the evolution that is generated by such a Hamiltonian both in the classical
and quantum settings is exactly the same. On the other hand, the Hamiltonians
that are linear in one of the basic variables, e.g. in $q$, generate the same evolution
for the infinite set of variables $(q,p,G^{n,0},G^{n,1})$ as for their classical counterparts
$(q,p,C^{n,0},C^{n,1})$ for all integer $n$.
In particular, the cosmological model that will be considered on this paper is described
by a Hamiltonian linear in $q$. In addition, when the cosmological constant is vanishing
this Hamiltonian will turn out to be linear in both $q$ and $p$. Thus, for that case,
the Hamiltonian will be harmonic and the quantum and classical (distributional)
evolutions it generates will be completely indistinguishable.

Finally, let us comment that the stationary states can also be considered within the present formalism.
The stationary states correspond to fixed points of the dynamical system under consideration and
thus its corresponding moments can be obtained by solving the algebraic system of equations
obtained by dropping all time derivatives from the equations of motion for $(q,p, G^{a,b})$.
In particular, following this procedure, the moments of the classical and quantum stationary states
of the harmonic and the quartic oscillators were studied in \cite{Bri14b}.

\section{Application to a cosmological model}
\label{sec_cosmology}

\subsection{The classical cosmological model with positive cosmological constant}

\begin{figure}
\includegraphics{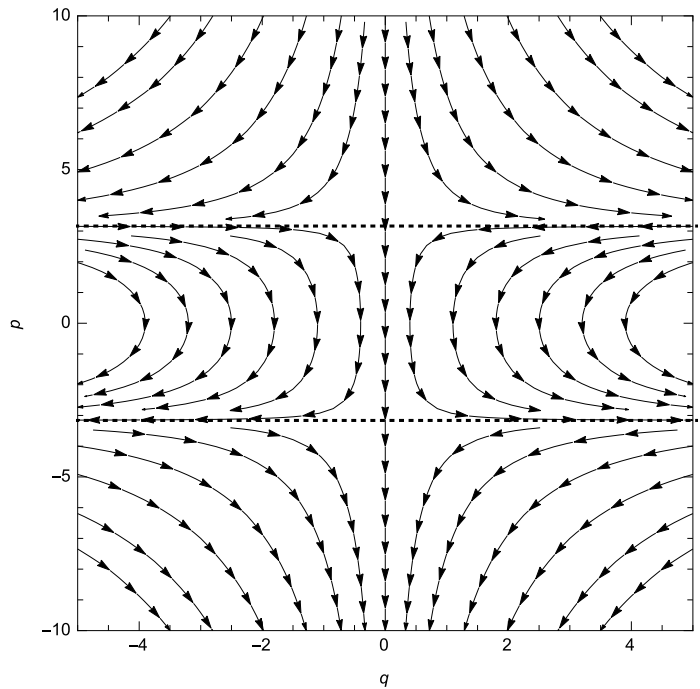}
\caption{In this plot the phase space corresponding to the cosmological model
under consideration is shown for the value $\Lambda=10$. The dotted lines stand
for the curves $p=\pm\sqrt{\Lambda}$, which can not be crossed by any orbit.
These lines divide the phase space in three different regions. The upper region
corresponds to expanding solutions. The lower region is the time reversed of
the latter one and thus describes collapsing orbits. Finally, the intermediate
region corresponds to bouncing solutions.}\label{phasespace}
\end{figure}

Let us assume a cosmological model of a homogeneous, isotropic and spatially flat
universe with a massless scalar field $\phi$ as matter content and positive
cosmological constant $\Lambda$. As usual in general relativity, this system is described by a
Hamiltonian constraint, as opposed to a physical Hamiltonian.
Nonetheless, it is possible to deparametrize the system and use the conjugate
momentum of the scalar field $p_\phi$, which is a constant of motion, as a
physical Hamiltonian.
The Friedmann equation corresponding to this system reads
\begin{equation}\label{friedmann}
\left(\frac{a^\prime}{a}\right)^2= \frac{4\pi G}{3}\frac{p_{\phi}^2}{a^6}+\Lambda,
\end{equation}
where $a$ is the scale factor and the prime stands for derivative with respect
to the cosmic time. By choosing Newton constant as $\frac{4\pi G}{3}=1$
for convenience, it is straightforward to solve this equation for
$p_\phi$ and define our physical Hamiltonian as,
\begin{equation}\label{h_pphi}
H:=p_{\phi}=a^2 \sqrt{|a^{\prime 2}-\Lambda a^2|},
\end{equation}
where the absolute value has been taken to extend this Hamiltonian
to the region $a^{\prime 2}<\Lambda a^2$.
In this way, the scalar field $\phi$ will play the role of time. Nevertheless,
this Hamiltonian must still be written in terms of the canonical variables.
Such variables are directly related to the scale factor of
the universe as $q=(1-x)a^{2-2x}$ and $p=-a^{2x}a'$. Then, the physical Hamiltonian
takes the following form
\begin{equation}
H=(1-x)q\sqrt{|p^2-\Lambda[(1-x)q]^{(1+2x)/(1-x)}|}.
\end{equation}
The parameter $x$ characterizes
different possible cases of lattice refinement of an underlying discrete state,
characteristic of a loop quantization \cite{Boj06, Boj08}. A
value around $x=-1/2$ seems to be favored by several independent
phenomenological and stability analysis in the context of loop quantum cosmology
\cite{APS06, BCK07, BH08, NeSa07, BCK12}.
Nevertheless, in the (geometrodynamical) quantization studied here,
with the usual canonical commutation relation, in principle this parameter does not play any role.
Thus, as was done in Ref. \cite{BBH11}, $x=-1/2$ will be chosen since it turns out
to be very convenient because it leaves the Hamiltonian as a linear function of the position:
\begin{equation}
H=\frac{3}{2}q\sqrt{|p^2-\Lambda|}.
\end{equation}
The Hamilton equations take then the following form,
\begin{eqnarray}
\dot{q}&=& \frac{3}{2} q\,p\frac{sg(p^2-\Lambda)}{\sqrt{|p^2-\Lambda|}},\label{eq1}\\
\dot{p}&=&-\frac{3}{2} \sqrt{|p^2-\Lambda|},\label{eq2}
\end{eqnarray}
where $sg$ is the sign function and the dot stands for derivative with respect to $\phi$.
The solution to these equations can be found analytically \cite{BBH11} but here, in order to
complement the discussion of that reference and to show in a more intuitive way the
dynamics of this system in terms of the chosen variables, the corresponding
phase space diagram is shown in Fig. \ref{phasespace}.
With the chosen value of the parameter $x$, $q$ is proportional to the volume of the universe $a^3$,
and one would then naturally define it as positive definite.
Nonetheless, for illustrational purposes, both negative and positive
values of the position $q$ have been plotted. In fact, as can be seen in the
diagram, the phase space is symmetric under a change of sign in $q$ (as the
whole system is symmetric under a change of sign of $a$). For positive $q$ (as well
as for negative $q$), the
phase space contains three disjoint regions separated by the lines $p=\pm\sqrt{\Lambda}$.
The upper and lower sectors correspond to $p^2>\Lambda$,
whereas in the middle sector $p^2<\Lambda$. This latter region
is the one that in principle is not allowed by the Friedmann equation (\ref{friedmann})
(the momentum of the scalar field would need to be imaginary)
but has been constructed extending the Hamiltonian by
taking the absolute value inside the square root (\ref{h_pphi}).

The explicit equation for
the orbits can be obtained analytically,
\begin{equation}
(p^2-\Lambda)q^2=k,
\end{equation}
with $k$ an integration constant. Note that positive values of the constant $k$
correspond to the upper and lower sectors, whereas negative values stand
for orbits in the middle region. Between those solutions, there is also
the degenerate solution $q=0$, that corresponds to a universe
with zero volume.

In the upper region all orbits begin at the positive infinite of $p$ and
vanishing $q$, with a value $\phi\rightarrow-\infty$ of the affine parameter,
and reach the asymptote $p=\sqrt{\Lambda}$ at an
infinite value of $q$ but at finite value of the affine parameter $\phi=\phi_{\rm div}$.
We will be interested in this region that corresponds to an expansion of
the universe. The lower region is the time reversed of this latter one.
There the system begins with an infinite volume and collapses.
In fact, every orbit of this region is the analytic extension of the orbit
with the same integration constant $k$ of the upper region for values
$\phi\in(\phi_{\rm div},\infty)$ of the affine parameter \cite{PaAs12}.
On the other hand, the middle region is nonphysical and corresponds to bouncing
solutions that begin with infinite volume, collapse until a minimum value
of $q=\sqrt{|k|/\Lambda}$ with $p=0$ and then expand again. Since we
will be working just in the upper sector, from here on, the absolute values
that appear in the Hamiltonian, as well as in the equations of motion, will
be removed without loss of generality.

\subsection{Evolution equations for classical and quantum moments}\label{sec_cosmologymoments}

Following the procedure described in Sec. \ref{sec_formalism}, one can construct
the effective Hamiltonian $H_Q$ for this system as \cite{BBH11}
\begin{eqnarray}\label{effectivehamiltonian}
H_Q&=& \frac{3}{2}q\sqrt{p^2-\Lambda}+ \frac{3}{2}\sqrt{\Lambda}
\sum_{n=2}^{\infty} \frac{\Lambda^{-n/2}}{n!}\bigg[ q\, T_{n}(p/\sqrt{\Lambda}) G^{n,0}+ \,n\,\sqrt{\Lambda}\, T_{n-1}(p/\sqrt{\Lambda}) G^{n-1,1}\bigg]\,,
\end{eqnarray}
where the function
\begin{equation}
 T_n(x):= \frac{\md^n}{\md x^n} \sqrt{x^2-1}
\end{equation}
has been defined. Making use of this Hamiltonian
it is straightforward to obtain
the infinite system of equations
that rules the evolution of the expectation values $(q,p)$ and quantum moments
$G^{a,b} $,
\begin{eqnarray}\label{dotq}
\dot{q} &=& \frac{3}{2}q\frac{p}{\sqrt{p^2-\Lambda}} + \frac{3}{2}
\sum_{n=2}^{\infty} \frac{\Lambda^{-n/2}}{n!}\bigg[ q\, T_{n+1}(p/\sqrt{\Lambda}) G^{n,0}
+\, n \,\sqrt{\Lambda}\, T_{n}(p/\sqrt{\Lambda}) G^{n-1,1}\bigg]\,,\\\label{dotp}
\dot{p}
&=& -\frac{3}{2}\sqrt{p^2-\Lambda}- \frac{3}{2}\sqrt{\Lambda}
\sum_{n=2}^{\infty} \frac{\Lambda^{-n/2}}{n!}
T_n(p/\sqrt{\Lambda}) G^{n,0},\\\label{dotG}
\dot{G}^{a,b} &=&\frac{3}{2}\sqrt{\Lambda}
\sum_{n=2}^{\infty} \frac{\Lambda^{-n/2}}{n!}\bigg[ q\, T_{n}(p/\sqrt{\Lambda}) \left\{ G^{a,b},G^{n,0} \right\}
+n\sqrt{\Lambda} T_{n-1}(p/\sqrt{\Lambda}) \left\{ G^{a,b},G^{n-1,1} \right\} \bigg],
\end{eqnarray}
where the Poisson brackets between moments have been left indicated.
(For explicit expressions of these brackets the reader is referred to
\cite{BBH11, Bri14}). The evolution equations for classical moments and
expectation values $(q,p,C^{a,b})$ are obtained from these previous
ones just by imposing $\hbar=0$. Such
$\hbar$ terms appear when computing the Poisson brackets between moments.

Sometimes the meaning is not clear from the context thus, following
the notation of \cite{Bri14b},
the solution of the quantum system (\ref{dotq}--\ref{dotG})
will be denoted as $q_q(\phi)$. On the other hand, the
solution of the classical distributional system [that is, the one
obtained from (\ref{dotq}--\ref{dotG}) by replacing all $G^{a,b}$
by it corresponding $C^{a,b}$ and imposing $\hbar=0$]
will be denoted by $q_c(\phi)$. Finally, the classical point
trajectory [the solution to Eqs. (\ref{dotq}--\ref{dotp})
dropping all moments] will be referred as $q_{class}(\phi)$.
The same notation is used for the variable $p$.

The system we are dealing with has a very particular Hamiltonian
since it is linear on the position variable $q$. That is why only moments of the
form $G^{n,0}$ and $G^{n,1}$ appear in the effective Hamiltonian
(\ref{effectivehamiltonian}), and in the second entry of the
Poisson brackets of Eq. (\ref{dotG}), as well as in the
equations for the expectation values (\ref{dotq}--\ref{dotp}). It can be shown that,
for such a system,
the infinite set of variables ($q_q$, $p_q$, $G^{n,0}$, $G^{n,1}$) for all integer $n$
obey a close system of equations, decoupled from the rest of the moments,
which does not contain any $\hbar$ term \cite{Bri14}. Therefore, given the same initial
data, the expectation values $q_q$, $p_q$, as well as
all moments of the form $G^{n,0}$ and $G^{n,1}$ follow exactly the same evolution
as their classical distributional counterparts ($q_c$, $p_c$, $C^{n,0}$, $C^{n,1}$).
This is completely generic for any initial data. Therefore, for this kind of Hamiltonians
$q_c(\phi)=q_q(\phi)$ and $p_c(\phi)=p_q(\phi)$ for all times. However, this does not mean that they follow the classical point
trajectory, $q_{class}(\phi)\neq q_c(\phi)$, since moment terms appear in Eqs. (\ref{dotq}--\ref{dotp}).
In the following subsections
different initial data will be considered in order to compare the classical and quantum
evolution of this cosmological model. And, thus, in order to find any difference,
moments not contained in that subset will have to be checked.

Finally, note that the $\Lambda=0$ is a harmonic case since the Hamiltonian
is linear in both position and momentum variables. This harmonic case has
very special properties. In particular, all orders are decoupled, and
there is no
$\hbar$ in any of the evolution equation of the moments. Thus
classical and quantum moments obey exactly the same set of equations.
In fact, it is easy to find the analytic solution for all variables,
\begin{eqnarray}
q(\phi)&=&q_0 \exp{\left[\frac{3}{2} (\phi-\phi_0)\right]},\\
p(\phi)&=&p_0 \exp{\left[-\frac{3}{2} (\phi-\phi_0)\right]},\\
G^{a,b}(\phi)&=&G^{a,b}_0 \exp{\left[\frac{3}{2} (b-a) (\phi-\phi_0)\right]}
\label{Gabsolharmonic}\,,
\end{eqnarray}
$(q_0,p_0,G^{a,b}_0)$ being the value of each function at $\phi=\phi_0$.

\section{Numerical implementation}\label{sec_numerical}

\subsection{Initial data}

In order to extract physical information from the infinite set of equations
for quantum moments (\ref{dotq}--\ref{dotG}), as well as for its corresponding classical counterpart
system, it is necessary to resort to numerical methods.
In addition, for practical purposes, one needs to consider a cutoff, that is, a maximum
number $N$ for which all moments of an order greater than $N$ are dropped.
This fact shows that this method of moments is well suited for
peaked states, when high-order moments are negligible. Nonetheless, the usual
tendency of the dynamics is to spread out the state so that, from certain point on,
this method will not give trustable results. There are several control methods
to know when this happens. On the one hand, one should study the convergence
of the solution with the cutoff. This is done by solving the system of equations
with different cutoffs and checking that the difference between solutions with
consecutive cutoffs tends to zero. On the other hand, the conservation of constants
of motion should also be taken into account. In the present model the Hamiltonian
$H_Q$ (and $H_C$ for the classical system) is conserved. Finally, the high-order
inequalities obtained in \cite{Bri14} should also be fulfilled during the whole evolution.
All these control methods have been used to test the numerical implementations
that will be presented below.

The goal of the present paper is to compare the evolution of the quantum
and classical (distributional) systems within the context of the formalism
of moments described above.
With that purpose, similar numerical evolutions to those that were performed
in Ref. \cite{BBH11} will be performed but in this case not only for
quantum moments, also for their classical counterparts. In that reference
a Gaussian state was chosen as initial state. The moments corresponding
to such a state read,
\begin{eqnarray}\label{gaussian_moments}
G^{a,b}=
\left\{
\begin{array}{c}
2^{-(a+b)} \hbar^a\sigma^{b-a}\frac{a!\,b!}{\left(\frac{a}{2}\right)!\left(\frac{b}{2}\right)!}
\quad{\rm if}\, a \,{\rm and}\, b\, {\rm are}\,{\rm even},\\
\\
 0\quad\quad\quad\quad\quad\quad{\rm otherwise},
\end{array}
\right.
\end{eqnarray}
$\sigma$ being the width of the Gaussian. As can be seen,
this state is of a very special kind, as many of its moments vanish.
In fact, in order to check whether the results obtained depend
on the fact that these moments are vanishing; apart
from the Gaussian initial state, here another two initial states
will also be considered. In other words, three evolutions will be performed.
The first one with a Gaussian state for both classical
and quantum moments. Whereas the initial state for the second
evolution will be given by evolving the Gaussian state with the
quantum equations during a short period of time ($0.1\phi_{\rm div}$).
At this point all moments will have been excited and the resulting
state will be a slightly deformed Gaussian,
which will be used as initial state for both classical and quantum
systems. Obviously, both evolutions will give the same result
for the quantum moments but not for the classical ones.
Finally, the initial data for the third evolution will be given
by all classical and quantum moments taking the nonvanishing value
\begin{equation}\label{nongaussian_moments}
G^{a,b}=\,a!\,b!\,\hbar^{\frac{a+b}{2}},
\end{equation}
which is allowed by all high-order uncertainty relations obtained in \cite{Bri14}.
This is a distribution quite different from the Gaussian
(\ref{gaussian_moments}), which will be used to check
the generality of the results obtained in the previous two cases.

Regarding more technical issues, the numerical evolutions have
been performed for all cutoffs from $N=2$ to $N=10$ in order
to verify the convergence of the method with the cutoff order.
Furthermore, different values of the cosmological constant
have been considered: small ($\Lambda=1$), intermedium ($\Lambda=10^4$),
and large ($\Lambda=9\times 10^7$). The initial conditions for the position and
the momentum have been chosen as $p(0)=10^5$, $q(0)=1$
to ensure that the state is on the upper region of the phase space
shown in Fig. \ref{phasespace} and corresponds to an expanding
solution. In addition, the Gaussian width has been taken as $\sigma=\sqrt{\hbar}$
so that the expression for a moment $G^{a,b}$ is completely symmetric
on its indices $a$ and $b$. In this way, initially $G^{a,b}=G^{b,a}$ and
the fluctuation (as well as higher-order moments) of the position are
equal to those of the momentum. Finally, the numerical value of the Planck
constant has been chosen as $\hbar=10^{-2}$.

In summary, the main difference with respect to the Gaussian state
used in Ref. \cite{BBH11} is that its width was chosen as $\sigma=10^{-2}$,
that is, smaller than in our present case with $\sigma=10^{-1}$. In
addition, in that reference, $\hbar$ was chosen to be
equal to one in order to make the backreaction effects more clearly
visible. However, qualitatively the results do not change with such
modifications.
Here, in order to analyze the purely quantum effects,
which appear in the quantum equations of motion as a power series
in $\hbar^{2}$, it is necessary to consider a smaller Planck constant,
otherwise all terms of the form $\hbar^{2n}$ would be of the same order.

In the next subsection we will comment, mainly qualitatively,
the behavior of different variables throughout evolution for
the Gaussian initial state (\ref{gaussian_moments}) for both classical
and quantum moments;
whereas in Subsec. \ref{sec_quantitative} the differences between classical and quantum
moments will be analyzed quantitatively in detail. 

\subsection{Description of the evolution}

\begin{figure}
\begin{minipage}{0.47\textwidth}
\centering
\includegraphics[width=\textwidth]{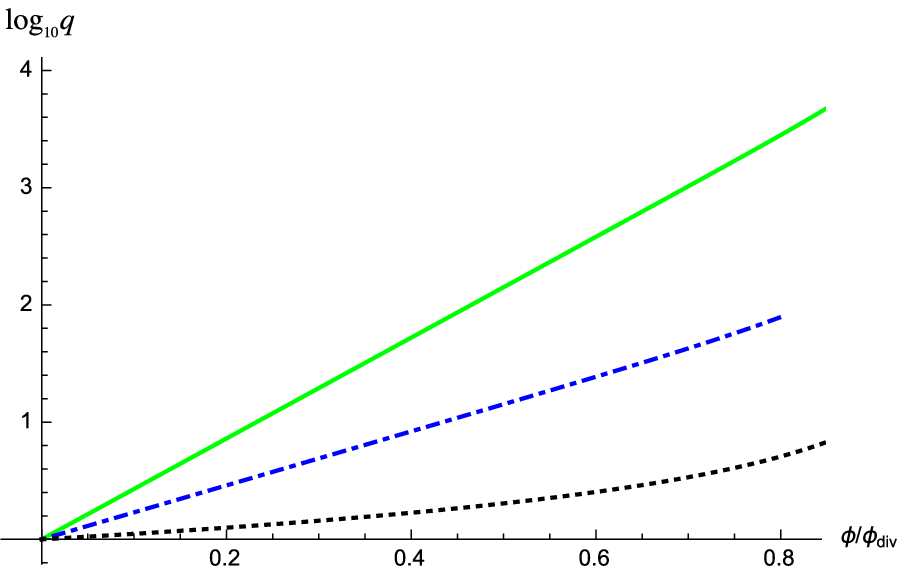}
\caption{In this figure the evolution of the volume $q$ is plotted
for three different values of the cosmological constant on a logarithmic
scale. The green (continuous) line stands for $\Lambda=1$,
the blue (dot-dashed) line corresponds to $\Lambda=10^4$,
whereas the black (dotted) line represents the solution with  $\Lambda=9\times10^7$.
Note that for the same value of $\phi/\phi_{\rm div}$,
$q(\phi/\phi_{\rm div})$ is larger, the smaller the
value of the cosmological constant.}\label{fig_volume}
\end{minipage}
\begin{minipage}{.08\textwidth}
\end{minipage}
\begin{minipage}{.49\textwidth}
\centering
\includegraphics[width=\textwidth]{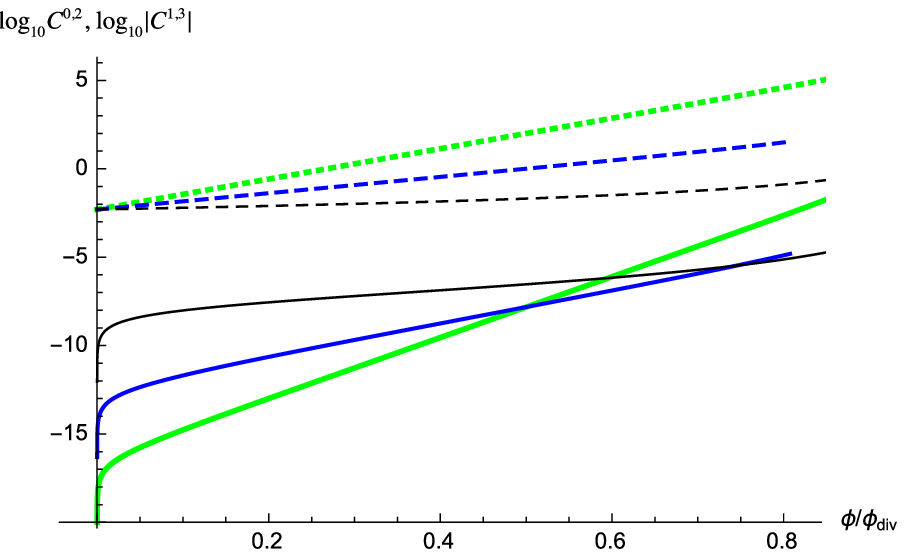}
\caption{In this plot the evolution of the moments $C^{0,2}$
and $C^{1,3}$ is shown in a logarithmic scale for three
different values of the cosmological constant and for an initial Gaussian state. Continuous lines
correspond to $C^{1,3}$, which is an initially vanishing moment,
whereas dashed lines stand for $C^{0,2}$. The color (thickness)
of the lines represent different values of $\Lambda$:
green (thickest) for $\Lambda=1$, blue (intermedium)
for $\Lambda=10^4$, and black (thinnest) for $\Lambda=9\times 10^7$.
Note that the larger the value of $\Lambda$, the larger the slope
of the moment in terms of $\phi/\phi_{\rm div}$.}\label{fig_momentsevolution}
\end{minipage}
\end{figure}

Regarding the expectation values,
as has already been commented in the preceding section, due to the
linearity of the Hamiltonian they have exactly the
same evolution both for the classical (distributional) case as for
the quantum case, that is, $q_q=q_c$ and $p_q=p_c$ at all times.
Nonetheless, the backreaction on the classical orbits is not
vanishing due to the presence of moments in Eqs. (\ref{dotq}, \ref{dotp}).
In other words, the centroid of a (classical or quantum) distribution
will not follow a classical point trajectory on phase space.
This backreaction can be measured by defining the differences
$\delta q:=q_q-q_{class}$ and $\delta p:=p_q-p_{class}$.
[Note that in principle $q_q(t)$ should be the solution of Eq. (\ref{dotq})
by considering the infinite set of equations. We obtain an approximate
version of this solution by imposing the cutoff $N=10$.]
Regarding this backreaction on the classical orbits,
though less severe due to the smaller numerical value of $\hbar$,
we obtain similar results as in Ref. \cite{BBH11}.
In particular the deviation from the classical orbit is greater
as we move to larger values of the cosmological constant.
More precisely, measured at $0.8\phi_{\rm div}$, the relative change
in volume is $\delta q/q\approx10^{-7}$ for the large cosmological
constant case. In the other two cases, due to its smallness,
this effect is mixed
with numerical error and it turns out to be difficult to measure
but, in any case, we have that $\delta q/q\lessapprox 10^{-10}$ throughout
evolution. In all cases there is enhancement of the divergence in
the sense that the position $q_q$ (or $q_c$) corresponding to
the centroid of a probability distribution approaches the divergence
faster than the classical point orbit $q_{class}$.
On the other hand, the deviation (in absolute value) of the momentum $p$
is not that pronounced and for all cases, during the whole evolution until the mentioned
time, $\delta p/p\lessapprox10^{-10}$. Thus, in this sense, the
case with larger cosmological constant $\Lambda=9\times 10^7$
is the one where the backreaction effects are more relevant.
The evolution of the volume $q$ is represented in Fig. \ref{fig_volume}
for different values of the cosmological constant.

Regarding the evolution of different moments, as already commented
above, due to the linearity of the Hamiltonian,
pure fluctuations of the momentum $G^{n,0}$ and moments
of the form $G^{n,1}$ coincide with their classical counterparts:
$G^{n,0}=C^{n,0}$ and  $G^{n,1}=C^{n,1}$ for all times. For the rest of the moments
it can be asserted that,
the behavior of classical moments $C^{a,b}$ is qualitatively similar
to their quantum counterparts $G^{a,b}$ except for those
of the form $G^{0,2n+1}$. In the next subsection we will
analyze in more detail the differences between both classical
and quantum moments, and particularly
the corresponding to these pure-odd fluctuations of the position $G^{0,2n+1}$.
For the rest of this subsection, and unless explicitly stated,
all that is said about the classical moments $C^{a,b}$ applies
equally to the quantum ones $G^{a,b}$.
In summary, the generic behavior of the moments that will be commented
below can be seen in Fig. \ref{fig_momentsevolution} for the particular case
of two moments: $C^{0,2}$ and $C^{1,3}$.

For a Gaussian state all moments with an odd index are vanishing.
These moments get excited as soon as the evolution begins because their time derivative is nonzero.
After that excitation generally moments $|C^{a,b}(\phi)|$ with $a> b$, including those that have been excited from an initial vanishing value,
behave as an (approximately exponentially) decreasing function; whereas
moments with $a\leq b$ increase in absolute value.
This behavior is inherited from the solution (\ref{Gabsolharmonic}) for the harmonic case $\Lambda=0$.
[Nevertheless, even if it is very convenient to have such a picture, several moments
break this rule (usually moments that are supposed to be decreasing turning out to be increasing
with time), specially as the value of the cosmological constant is larger.] In fact, this general tendency can
be intuitively expected by looking at the phase space of the system shown in Fig. \ref{phasespace}.
Let us assume that initially we have a homogeneous and compact probability distribution
with the shape of a circle centered at small $q_0$ and large $p_0$. When the evolution begins,
each point of that circle follows its corresponding orbit and the circle will get deformed.
In particular, as these points approach the divergence at $p=\sqrt{\Lambda}$, the initial
circle will be enlarged in the horizontal ($q$) direction whereas it will be contracted
in the vertical ($p$) direction. Thus, the state will be more spread in the $q$ direction
but more peaked in the $p$ direction. This is exactly what we have described in
terms of moments: moments $C^{a,b}$ with more weight in the $p$ direction ($a>b$) will be decreasing, and the others increasing, in absolute value.

On the other hand, regarding the sign of the moments, during the initial stages
of the evolution we observe that,
\begin{eqnarray}
C^{\rm even, even}>0,\,\,\quad\quad C^{\rm even, odd}>0,\\
\mbox{}\quad C^{\rm odd, even}<0,\,\,\quad\quad C^{\rm odd, odd}<0.
\end{eqnarray}
The only exception to these rules are the pure-odd fluctuations of the momentum $C^{{\rm odd},0}$,
which are positive. The same applies to quantum moments, except for those
of the form $G^{0,{\rm odd}}$ that, as will be explained below, are initially excited to a negative
value and then change sign during the evolution. Note that the first of the inequalities, which states that moments
with both even indices must be positive, is implied by the very definition of the moments
and must be obeyed at all times for any system \cite{Bri14}.

Even if the commented qualitative features for the moments are
independent of the value of the cosmological constant, there is indeed
some dependence. In particular, the larger $\Lambda$ the lower the slope,
in terms of $\phi/\phi_{\rm div}$, of a given moment is.
[Note, however, that $\phi_{\rm div}$ also depends
on the cosmological constant since $\phi_{\rm div}:=\phi_0+2/3 \tanh^{-1}(\sqrt{p_0^2-\Lambda}/p_0)$].
In addition, the excitation value of the initially vanishing moments
increase, in absolute value, as we consider larger values of the cosmological constant.
However, since the slope of the moments is larger for small $\Lambda$,
at the end stages of the evolution an increasing (decreasing) moment
$|C^{a,b}|$ has usually a larger (smaller) value for smaller values of
$\Lambda$, as can be seen in the example of Fig. \ref{fig_momentsevolution}.
Even so, the relative fluctuations $C^{a,b}/(q^a p^b)$ are generally
larger the larger the value of the cosmological constant.

\subsection{Quantitative comparison between classical
and quantum moments}\label{sec_quantitative}

Following the notation of \cite {Bri14b}, we introduce the following operators,
$\delta_1$ and $\delta_2$,
to quantitatively measure the difference between the classical and quantum
evolution of the present system:
\begin{eqnarray}
\delta_1 q(\phi)&:=&q_c(\phi)-q_{class}(\phi),\\
\delta_2 q(\phi)&:=&q_q(\phi)-q_c(\phi).
\end{eqnarray}
Note that the action of $\delta_1$ on a given
moment $G^{a,b}$, $\delta_1G^{a,b}$, is not
defined since there are no moments in the classical point orbit;
whereas $\delta_2 G^{a,b}:=G^{a,b}-C^{a,b}$. The first operator $\delta_1$
can be understood as a measure of distributional effects, whereas the
second one $\delta_1$ measures the strength of purely quantum effects.

Due to the properties of linear Hamiltonians,
in this case $\delta_2 q$ and $\delta_2 p$ are vanishing. Therefore, the departure $(\delta_1q,\delta_1p)$
of the centroid from the classical (point) trajectory on the phase space is
uniquely due to distributional effects, which are indeed exactly reproduced
by a similar distribution evolving on the classical phase space.
Furthermore, $\delta_2G^{n,0}$ and $\delta_2G^{n,1}$ are also vanishing
for any integer $n$ given any initial data.
Therefore, in order to check the relevance of the purely quantum effects
of this system, it is necessary to consider other kind of moments.

In order to check the robustness of certain results that will be commented below,
three different sets of initial data have been considered: a Gaussian state with moments given by (\ref{gaussian_moments}),
a slightly deformed Gaussian (by evolving the previous Gaussian with the quantum equations
during $0.1\phi_{\rm div}$), and a state with all nonvanishing moments
of the form (\ref{nongaussian_moments}). As one would expect, for the deformed Gaussian case,
the classical moments tend to show less divergence from their quantum counterparts as in the Gaussian case.

The analysis of the results of such numerical implementations has shown several
interesting features, which are listed below from $i/$ to $iv/$. Note that,
in order to remove possible spurious effect of the cutoff ($N=10$),
only moments up to order seven will be considered for the present discussion
since higher-order moments are the most sensitive ones to the effect of the cutoff.
Unless otherwise explicitly stated, the following results correspond to the Gaussian initial state.
In particular, we will refer to the implementations of the other two initial data at the final
part of item $iii/$.

$i/$ One of the important results is that all moments $G^{a,b}$,
except those of the form $G^{0,{\rm odd}}$, even if quantitatively different,
have the same qualitative behavior as their classical counterparts $C^{a,b}$.
Thus purely quantum effects act, as one would expect, as small perturbations by slightly deforming
the numerical values of different variables but not, in general, its qualitative behavior.

$ii/$ In absolute terms, $|\delta_2G^{a,b}|$ is largest for moments of the form
$G^{0,n}$ and $G^{1,2 n}$. This absolute difference tends to increase with
time and depends on the value of the cosmological constant. More precisely,
for the case with $\Lambda=1$ at a time $\phi=0.75\phi_{\rm div}$, 
the highest difference corresponds to the moment $G^{0,6}$ with $\delta_2G^{0,6}\approx6\times10^4$.
Such a large difference is only measured for this moment, which is also
the largest of all considered ones [$G^{0,6}(0.75 \phi_{\rm div})\approx 4\times10^{13}$];
the next highest value being $\delta_2G^{0,5}\approx -7$, whereas the rest of the moments
have an upper bound $|\delta_2G^{a,b}|< 0.1$. On the other hand, for the intermediate value of the
cosmological constant $\Lambda=10^4$ at the mentioned time $\phi=0.75\phi_{\rm div}$,
all absolute differences are $|\delta_2G^{a,b}|\leq 6\times 10^{-4}$,
whereas for large cosmological constant value $\Lambda=9\times 10^7$ this upper bound is much lower: $|\delta_2G^{a,b}|\leq 2\times 10^{-7}$.

$iii/$ In order to measure the strength of purely quantum effects in relative
terms, we define the relative difference $\delta_{a,b}:=\delta_2G^{a,b}/G^{a,b}$
and the ratio $r_{a,b}:=G^{a,b}/C^{a,b}$, which is probably more intuitive to
understand. These two objects are obviously related as $r_{a,b}=1-\delta_{a,b}$.

For the Gaussian initial state,
in relative terms, we find that moments $G^{a,b}$ with $a+b$ an odd number,
that is, with an even and an odd index, suffer the largest departure from their
classical counterparts $C^{a,b}$. [This statement, of course, exclude those
of the form $G^{n,0}$ and $G^{n,1}$.] For the remaining moments, even if
generically increasing (in absolute value) with time, so that the largest values are measured at
the end of the evolution, their relative changes are bounded as $|\delta_{a,b}|<4\times 10^{-10}$
for the cases $\Lambda=1, 10^4$; whereas for the case of a large cosmological
constant $\Lambda=9\times10^7$, $|\delta_{a,b}|<2\times 10^{-6}$.

\begin{figure}[h]
\begin{minipage}{.49\textwidth}
\centering
\includegraphics[width=\textwidth]{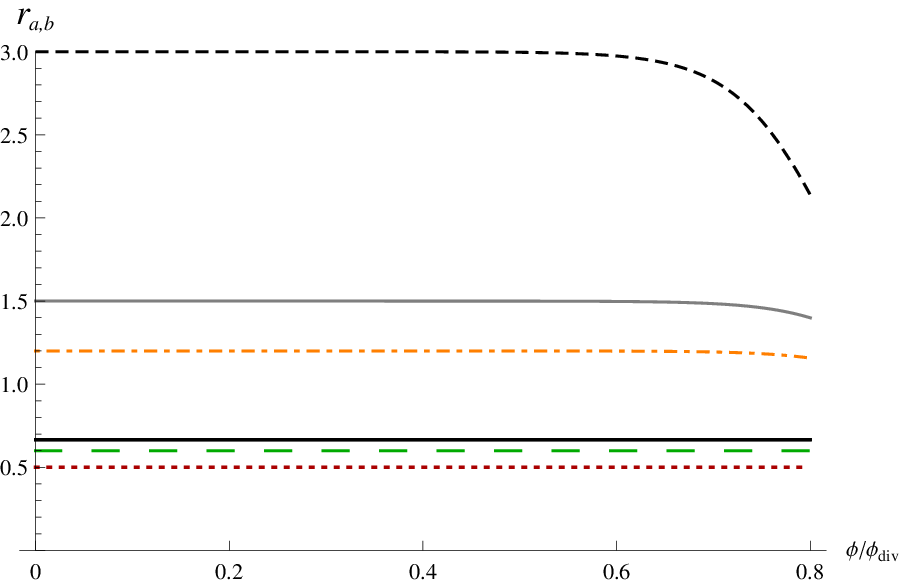}
\vspace{0.035cm}
\includegraphics[width=\textwidth]{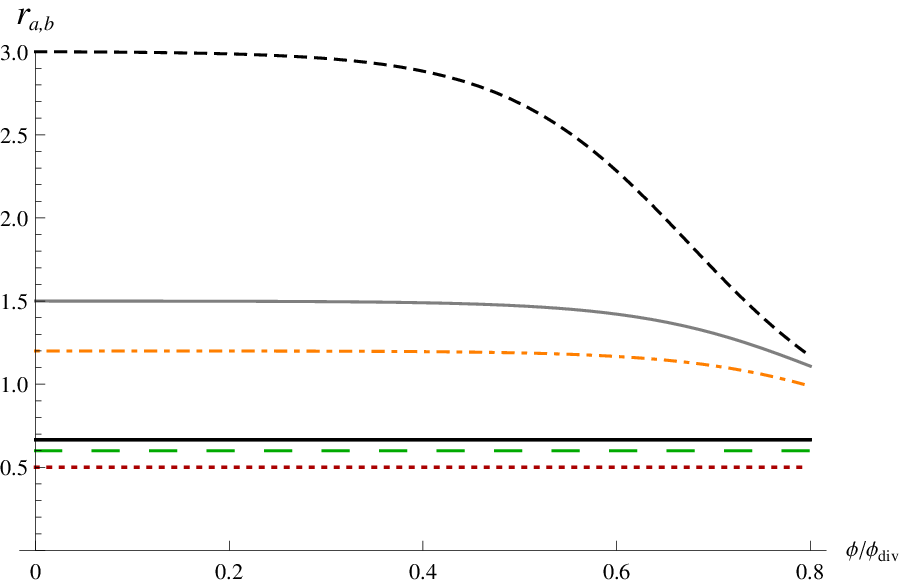}
\vspace{0.035cm}
\includegraphics[width=\textwidth]{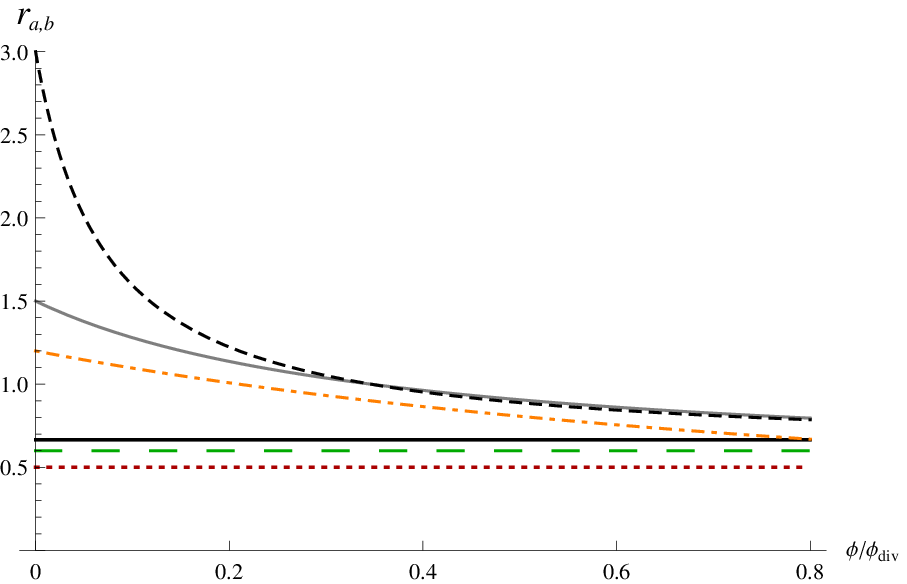}
\end{minipage}
\begin{minipage}{.49\textwidth}
\centering
\includegraphics[width=\textwidth]{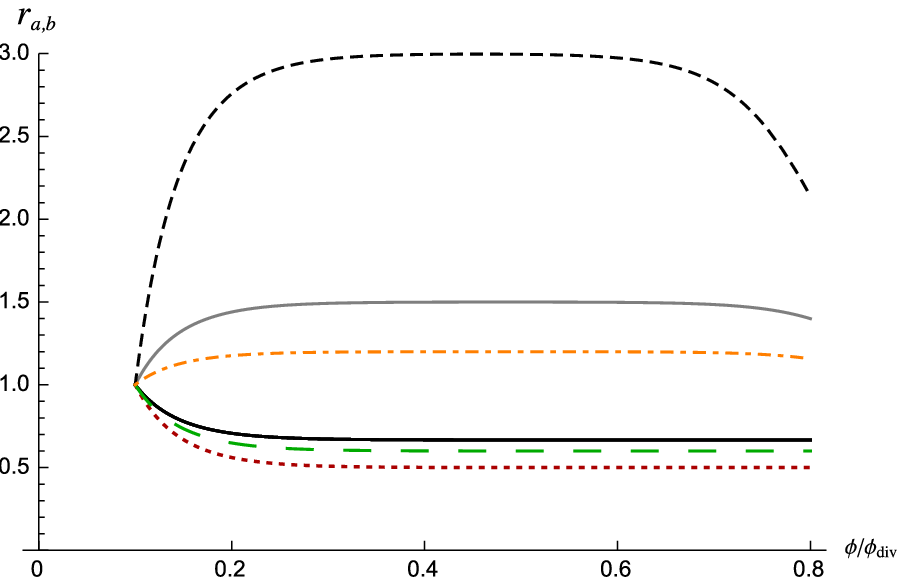}
\vspace{0.035cm}
\includegraphics[width=\textwidth]{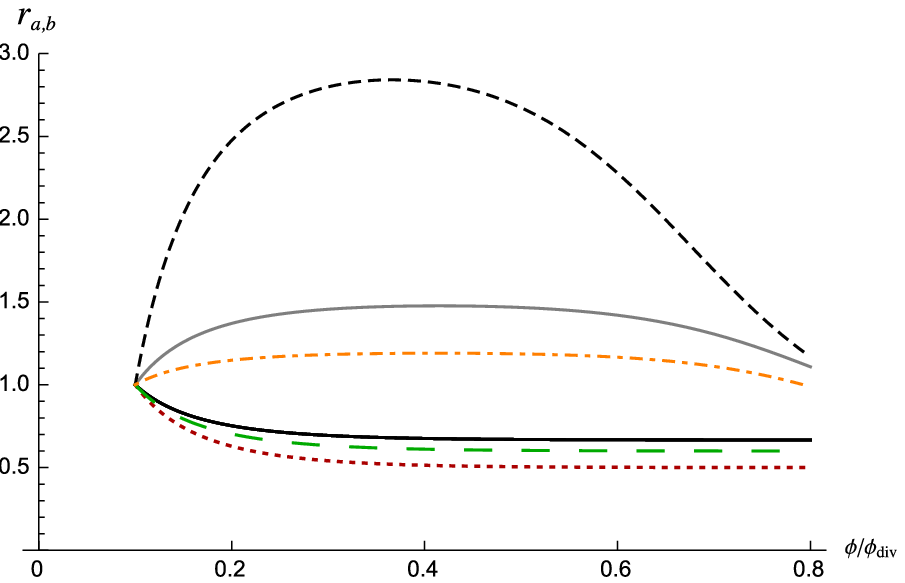}
\vspace{0.035cm}
\includegraphics[width=\textwidth]{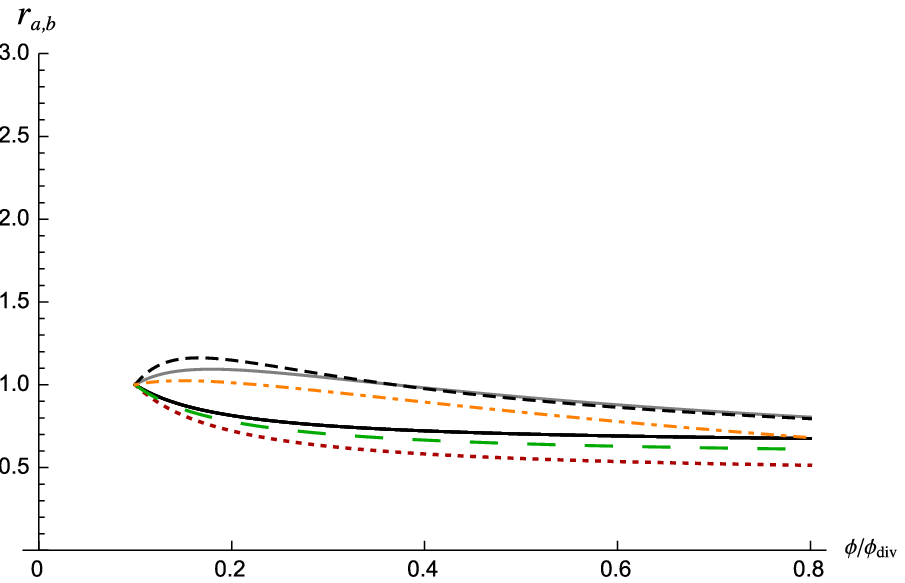}
\end{minipage}
\caption{In these plots the evolution of the ratios $r_{a,b}=G^{a,b}/C^{a,b}$
corresponding to quantum moments with highest deviation from their classical analogs,
excluding those of the form $G^{0,{\rm odd}}$, is shown. The left column corresponds
to the initial Gaussian state, whereas the right column stands for the initial deformed Gaussian case.
The upper plots correspond
to $\Lambda=1$, the medium ones to $\Lambda=10^4$, and the lower ones
to $\Lambda=9\times10^7$. The (continuous) black line represents the
three ratios $r_{1,2}$, $r_{1,4}$, and $r_{1,6}$. All of them
appear superposed since they follow very similar trajectories.
The ratios $r_{3,2}$ and $r_{3,4}$ correspond to the red (dotted)
and green (long-dashed) lines respectively.
Finally, $r_{2,3}$ appears
represented by the gray (continuous) line, $r_{2,5}$
by the black (dashed) line, and $r_{4,3}$
by the orange (dot-dashed) line.
Note that the late-time behavior is the same for both initial states.
}\label{ratios}
\end{figure}

Let us then focus on these moments with largest departure, that is, moments
$G^{a,b}$ with $a+b$ an odd number. The evolution of their ratios $r_{a,b}$
is plotted in Fig. \ref{ratios} for different values of the cosmological
constant and for both the Gaussian (left column) and the deformed Gaussian (right column) initial states.
For the Gaussian initial state their behavior can be separated in the following
three different kinds:
\begin{enumerate}
\item Moments of the form $G^{0,{\rm odd}}$. As already commented
above, a given moment of the form $C^{0,2n+1}$ in general does not follow, even
qualitatively, the same evolution as its quantum counterpart $G^{0,2n+1}$.
We will come back to these latter in the point $iv/$.
\item Moments of the form $G^{{\rm odd,even}}$. Remarkably, as can be seen in Fig. \ref{ratios}, the relative change
of these moments are constants of motion regardless the value
of the cosmological constant. The exception to this rule is $G^{5,2}$ which, for all values of the cosmological
constant, is initially excited to a value much higher than its classical
counterpart $C^{5,2}$. Thus their ratio $r_{5,2}$ is very small
but increases slowly with time. The ratios $r_{a,b}$ for the rest of the moments of this type
take approximately the following values:
\begin{eqnarray}
r_{1,2}\approx r_{1,4}\approx r_{1,6}\approx 2/3,\\
r_{3,2}\approx\ 1/2,\\
r_{3,4}\approx\ 2/5.
\end{eqnarray}
All these moments are negative and increasing (in absolute value)
throughout evolution.

\item  Moments of the form $G^{{\rm even,odd}}$. These moments, which, up to seventh order,
are just three ($G^{2,3}$, $G^{2,5}$, and $G^{4,3}$), are positive and increasing (in absolute value) during
evolution. They are initially excited to a higher value than their classical counterparts,
thus $r_{a,b}>1$. Remarkably, this initial value of the ratio is independent of the value of
the cosmological constant. Nonetheless, and contrary to the previous case,
the relative difference between classical and quantum moments does depend on time. More specifically,
ratios $r_{a,b}$ tend to decrease with time, which means that the classical moments increase faster than the quantum ones.
Furthermore, this decrease is faster the larger the value of the
cosmological constant. Interestingly, $r_{2,5}$ and $r_{2,3}$ seem
to tend to the same asymptote, as is more clearly visible in the last
plot of the left column of Fig. \ref{ratios}. For the case with largest cosmological constant ($\Lambda=9\times 10^7$),
the final values of all these ratios are lower than one, thus
the corresponding quantum moments $G^{a,b}$ end up being smaller
than their classical counterparts. On the other hand, for the other
two cases ($\Lambda=1$ and $\Lambda=10^4$), the final values of these
ratios are still larger than one.

\end{enumerate}
Note that all moments which have been analyzed in the last three
points were initially vanishing. Hence, we can divide the analysis
in two stages: the initial excitation and subsequent evolution.
All these moments are excited to different values
than their classical counterparts. Therefore, there is a clear
purely quantum effect acting on this excitation mechanism (that
is no other than the $\hbar$ terms present in the equations
of motion). Nevertheless, once this mechanism acts we observe
different behaviors. On the one hand, those moments that have
been initially excited to a smaller value than their classical
counterpart ($r_{a,b}<1$) keep their relative difference
constant throughout evolution, that is, they evolve in a very similar
way as their classical counterparts. This means that, for this kind of moments,
purely quantum effects act continuously during the whole evolution such
that $G^{a,b}$ is always proportional to $C^{a,b}$, with a constant
proportionality coefficient. On the other hand, the evolution of the quantum moments
that have been initially excited to a larger value than their classical counterparts
($r_{a,b}>1$) is slowed down by purely quantum effects. In such a way that
the corresponding classical moments increase faster (in absolute value).
Therefore, the net effect of the purely quantum terms is contrary in the initial excitation
of the quantum moment, when its value is risen with respect to its classical counterpart,
and during the subsequent evolution, when it is slowed down.

These results are quite unexpected, especially the constant behavior of
the ratio between certain moments. In fact, this was the main motivation
to consider other sets of initial data. Regarding the initial deformed Gaussian,
the results obtained are shown in the right-hand column of Fig. \ref{ratios}.
It is clearly seen that initially the ratios $r_{a,b}$ behave in a different
way as in the previous case, in particular none of them is kept constant.
Nonetheless, after a period of time, the classical moments follow the same
tendency they had for the Gaussian case and the final part of the evolution regarding the
different ratios have exactly the same shape as before.

In addition, just to test that this behavior is not completely generic,
as would be expected, for this particular issue another third set of initial data has been considered,
given by Eq. (\ref{nongaussian_moments}). In this case, the behavior
described above for the ratios $r_{a,b}$ of moments of the form
$G^{{\rm odd, even}}$ and $G^{{\rm even, odd}}$
disappear and the same tendency as the remaining moments is shown;
that is, all ratios are close to one initially and they depart
from one as the evolution advances. Nevertheless, this departure is
small and all relative differences are bounded as $|\delta_{a,b}|<5\times 10^{-8}$
for $\Lambda=1$, $|\delta_{a,b}|<4\times 10^{-7}$ for $\Lambda=10^4$,
whereas for the the large cosmological constant
$\Lambda=9\times10^7$ case $|\delta_{a,b}|<5\times 10^{-5}$.

Thus, we conclude that the behavior found for the Gaussian case is robust
under small deformation of the initial data, but not completely generic.
This situation, nonetheless, might be important if we know that
the initial state is an approximately semiclassical Gaussian state.

In any case, finding an analytical explanation of this surprising behavior is very difficult since
the system under consideration is a nonlinear and highly coupled set of differential equations.
As an example, in the appendix the purely quantum terms that appear in the equations
of motion for the moments $G^{2,2}$ and $G^{3,2}$ are shown. Even if the terms
are quite similar, in the sense that the same moments (except one) appear in both equations,
the behavior of their corresponding ratios $r_{2,2}$ and $r_{3,2}$, which somehow
quantifies the effect of the presence of these terms, are completely different.

\begin{figure}
\begin{minipage}{0.47\textwidth}
\centering
\includegraphics[width=\textwidth]{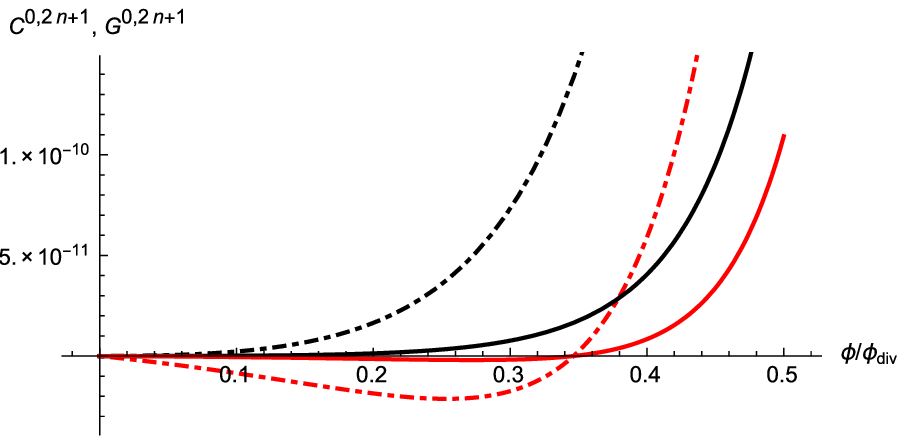}
\caption{The initial stages of the evolution of the moments $G^{0,3}$ (red dot-dashed line),
$G^{0,5}$ (red continuous line) and their classical counterparts $C^{0,3}$ (black dot-dashed line), $C^{0,5}$ (black continuous line) for the particular case with $\Lambda=9\times 10^7$.}
\label{G0305}
\end{minipage}
\begin{minipage}{0.47\textwidth}
\centering
\includegraphics[width=\textwidth]{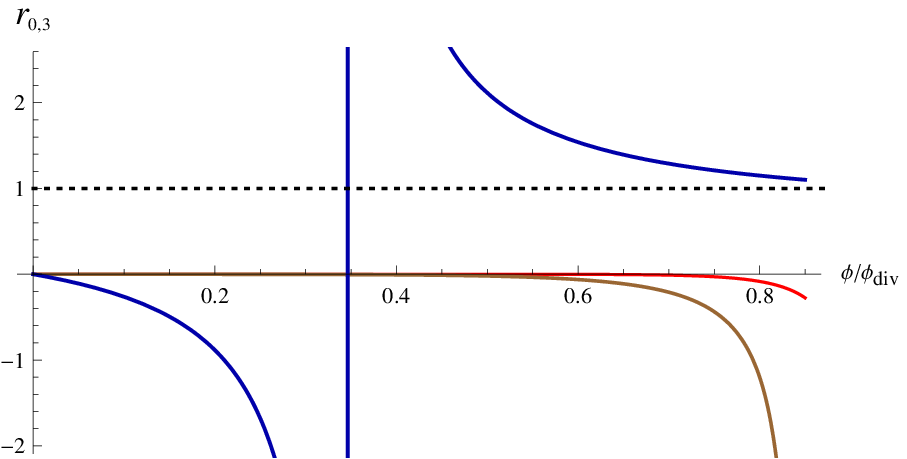}
\caption{The ratio $r_{0,3}$ is shown for different values of the cosmological
constant. The blue (thickest) line corresponds to $\Lambda=9\times 10^7$,
the brown to $\Lambda=10^4$, and the red (thinnest) line to $\Lambda=1$.
The dotted black line just represents the asymptote $r_{0,3}$=1. Note that,
for small and intermedium values of the cosmological constant, $r_{0,3}$
is negative throughout evolution due to the fact that $G^{0,3}$ is negative,
whereas $C^{0,3}$ is positive. On the contrary, in the case $\Lambda=9\times 10^7$,
$G^{0,3}$ flips sign, which produces the vertical line around $\phi=0.346\phi_{\rm div}$,
and the ratio $r_{0,3}$ tends to one asymptotically.}\label{r03}
\end{minipage}
\end{figure}

$iv/$ Finally, let us analyze the behavior of moments of the form $G^{0,{\rm odd}}$.
As has been commented above, these are the only moments that
follow qualitatively different trajectories in the classical and quantum settings.
Let us be more specific. For the Gaussian initial state in the classical case all moments $C^{0,{\rm{odd}}}$,
which are initially vanishing, are excited to certain positive value
and increase through evolution.  On the contrary, their quantum counterparts
$G^{0,{\rm odd}}$ are initially excited to a negative value (much smaller in absolute
value than their classical counterparts) and their evolution
is decreasing (that is, increasing in module), which makes their corresponding
ratio $r_{a,b}$ negative but close to zero. This decreasing behavior follows during
the whole evolution for the cases of small and intermedium cosmological
constant cases, which means that quantum moments increase faster (in absolute value)
than their classical counterparts. Nonetheless, for the large cosmological constant case $\Lambda=9\times 10^7$,
at certain point these moments get a minimum and increase afterwards. Remarkably at the
same time, around $\phi=0.346\phi_{\rm div}$, all moments $G^{0,{\rm odd}}$
cross zero and follow their growing behavior (see Fig. \ref{G0305}).
Finally, each quantum moment $G^{0,2n+1}$ tends asymptotically to its classical
counterpart $C^{0,2n+1}$,
as can be seen in Fig. \ref{r03}, which represents the evolution of the
ratio $r_{0,3}$ as an example of moments of this kind.
For the deformed Gaussian moments are initially negative. Nevertheless
all classical moments cross zero and tend to the above described evolution for the initial Gaussian case.

\section{Conclusions}\label{sec_conclusions}

In this paper the formalism presented in \cite{Bri14} to study the evolution of
classical and quantum probability distributions, by performing a decomposition on
its corresponding moments, has been applied to a particular cosmological model.
The goal was to study similarities and differences between the dynamics of
classical and quantum moments, in order to find physical consequences of purely
quantum terms, which are present in the equations of motion for quantum moments.

The cosmological model under consideration is a homogeneous
and isotropic universe with a massless scalar field and a positive cosmological
constant. The evolution of such a model is ruled by a Hamiltonian constraint,
which can be deparametrized by using the scalar field as internal time
and, thus, its conjugate momentum as the physical Hamiltonian. After introducing
an appropriate pair of conjugate variables, that represent the volume and the Hubble factor
of the universe, this physical Hamiltonian turns out to be linear in the volume $q$.

In fact,
it can be shown that
for any linear Hamiltonian in $q$ the evolution of the expectation values $(q,p)$,
as well as of the moments of the form $(G^{n,0},G^{n,1})$ for any integer $n$,
coincides both in the quantum and classical settings. In other words, given the same
initial conditions, the centroid and the mentioned moments of a classical
distribution evolve in exactly the same way as the corresponding variables
of a quantum distribution. Therefore, in order to find physical consequences
of the purely quantum effects, the dynamics of moments not contained
in that subset needs to be analyzed.

Nonetheless, for such a purpose,
it is necessary to resort to numerical methods
due to the complexity of the equations of motion. In particular, three different initial states
have been considered for both classical and quantum settings.
First a Gaussian in the volume has been chosen.
Second a slightly deformed Gaussian state
has been constructed by evolving the previous Gaussian state during a short period of time
with the quantum evolution equations. The obtained state has then be used as initial
condition for both classical and quantum moments.
Finally, the third set of initial
data is given by Eq. (\ref{nongaussian_moments}) and represents a state
which is not close to a Gaussian. This state has been mainly used to check
that the behavior of the ratios shown in Fig. \ref{ratios} is not completely generic.

The main result is that, in relative terms, moments of the form $G^{a,b}$ with $a+b$ an odd
number, are the ones that show most divergence from their classical analogs.
For the initial Gaussian state, all these moments are vanishing.
Since their time derivative is nonzero they are immediately excited as soon
as the evolution begins. This excitation value turns out to be quite different
for a given classical moment and its quantum counterpart. After this excitation,
two different behaviors have been observed. Quantum moments that have been excited to a smaller value
than their classical counterparts ($r_{a,b}<1$) evolve in a very similar
way as their classical analog, keeping the ratio $r_{a,b}$ constant.
On the other hand, evolution of the quantum moments that have been initially
excited to a larger value than their classical counterparts
($r_{a,b}>1$) is slowed down by purely quantum effects. In this way,
their corresponding classical moments grow faster in absolute value.
In summary, the effect of the purely quantum terms is contrary initially and during evolution.
While initially the value of the quantum moment is risen with respect to its classical analog,
during the subsequent evolution its growth is slowed down.

In addition,
for the initial deformed Gaussian state, even if initially different,
the ratios $r_{a,b}$ tend to the same behavior as explained above
for the Gaussian case.
On the contrary, the third set of initial data (\ref{nongaussian_moments}) does not follow this tendency.
In this latter case, all ratios depart only slightly from one.
Therefore, we conclude that 
this behavior of the different ratios is robust under small deformations of the initial
Gaussian state but, certainly, not completely generic.

\acknowledgments

The author thanks I\~naki Garay, Claus Kiefer, Manuel Kr\"amer, and Hannes Schenck
for discussions and comments. Special thanks to Martin Bojowald for interesting comments
on a previous version of this manuscript. Financial support from the Alexander von
Humboldt Foundation through a postdoctoral fellowship is gratefully acknowledged.
This work is supported in part by Projects IT592-13 of the Basque Government
and FIS2012-34379 of the Spanish Ministry of Economy and Competitiveness.

\appendix
\section{An example of purely quantum terms}

In this appendix, the purely quantum terms for the equations of the moments $G^{2,2}$
and  $G^{3,2}$ are shown. These are the terms with an explicit $\hbar$ factor that
appear in the right-hand side of their corresponding equations of motion:
\begin{eqnarray*}
\dot{G}^{2,2}&=&\frac{3 \hbar^2 \Lambda}{64 \left(p^2-\Lambda \right)^{17/2}}
\Big\{48 p \left(p^2-\Lambda \right)^6 G^{2,0}-24
   \left(\Lambda +4 p^2\right) \left(p^2-\Lambda \right)^5 G^{3,0}+40 p \left(3 \Lambda
   +4 p^2\right) \left(p^2-\Lambda \right)^4 G^{4,0}
  \\&-&30 \left(\Lambda ^2+8 p^4+12
   \Lambda  p^2\right) \left(p^2-\Lambda \right)^3 G^{5,0}+42 p \left(5 \Lambda ^2+8
   p^4+20 \Lambda  p^2\right) \left(p^2-\Lambda \right)^2 G^{6,0}
   \\&-&7 \left(5 \Lambda^3+64 p^6+240 \Lambda  p^4+120 \Lambda ^2 p^2\right) \left(p^2-\Lambda \right) G^{7,0}
   +9 p \left(35 \Lambda ^3+64 p^6+336 \Lambda  p^4+280 \Lambda ^2p^2\right)G^{8,0}\Big\}+\dots \\
\dot{G}^{3,2}&=&\frac{9 \hbar^2 \Lambda}{128 \left(p^2-\Lambda \right)^{17/2}}
\Big\{-16 \left(p^2-\Lambda \right)^7 G^{2,0}+48 p \left(p^2-\Lambda \right)^6G^{3,0}
-24 \left(\Lambda +4 p^2\right)\left(p^2-\Lambda \right)^5 G^{4,0}
\\&+&40 p \left(3 \Lambda +4 p^2\right)\left(p^2-\Lambda \right)^4 G^{5,0}
-30 \left(\Lambda ^2+8 p^4+12 \Lambda  p^2\right)\left(p^2-\Lambda \right)^3 G^{6,0}
+42 p \left(5 \Lambda ^2+8 p^4+20 \Lambda p^2\right) \left(p^2-\Lambda \right)^2 G^{7,0}
\\&-&7 \left(5 \Lambda ^3+64 p^6+240\Lambda  p^4+120 \Lambda ^2 p^2\right) \left(p^2-\Lambda \right) G^{8,0}
+9 p\left(35 \Lambda ^3+64 p^6+336 \Lambda  p^4+280 \Lambda ^2 p^2\right) G^{9,0}\Big\}+\dots,
\end{eqnarray*}
where dots stand for terms with no $\hbar$ factors. As can be appreciated, the purely quantum terms
that appear in these two equations are pretty similar, in the sense that the same moments appear (except one)
with similar polynomial coefficients depending on the expectation values $q$ and $p$, as well as
on the cosmological constant. Nonetheless, the effect of such terms are very different on each case.
On the one hand, $G^{2,2}$ is quite similar as $C^{2,2}$ during the whole evolution, keeping $r_{2,2}$
around one but increasing with time. On the other hand, for an initial Gaussian state, $r_{3,2}$ takes
a constant value around $1/2$, independently of the value of the cosmological constant. The behavior
of the latter ratio $r_{3,2}$ can also be seen in Fig. \ref{ratios} for the evolved Gaussian initial state.
In that case, it begins with a value of one, since both quantum and classical moments are initially equal
and decreases until reaching again the constant value of 1/2.

\end{document}